\documentclass[nohyper,12pt,letterpaper]{JHEP}
\usepackage[dvips]{epsfig}
\usepackage{amsfonts,amssymb}

\newcommand{\be}{\begin{equation}}
\newcommand{\ee}{\end{equation}}
\newcommand{\bea}{\begin{eqnarray}}
\newcommand{\eea}{\end{eqnarray}}
\newcommand{\ba}{\begin{array}}
\newcommand{\ea}{\end{array}}

\def\bbox{{\,\lower0.9pt\vbox{\hrule \hbox{\vrule height 0.2 cm
\hskip 0.2 cm \vrule height 0.2 cm}\hrule}\,}}
\newcommand{\dsl}{\pa \kern-0.5em /}

\newcommand{\nn}{\nonumber \\}

\newcommand{\EQ}{\begin{equation}}
\newcommand{\EN}{\end{equation}}

\renewcommand{\a}{\alpha}

\newcommand{\e}{\epsilon}
\def\bbox{{\,\lower0.9pt\vbox{\hrule \hbox{\vrule height 0.2 cm
\hskip 0.2 cm \vrule height 0.2 cm}\hrule}\,}}

\newcommand{\pa}{\partial}

\def\sac{\, , \qquad}
\newcommand{\eqn}[1]{(\ref{#1})}
\newcommand{\w}{\wedge}

\newcommand{\vp}{\varphi}

\newcommand{\cale}{{\cal E}}
\newcommand{\bcale}{{\bar{\cal E}}}
\newcommand{\calb}{{\cal B}}
\newcommand{\bcalb}{{\bar{\cal B}}}
\newcommand{\calf}{{\cal F}}
\newcommand{\call}{{\cal L}}

\newcommand{\calh}{{\cal H}}

\newcommand{\ione}{{\it 1}}
\newcommand{\itwo}{{\it 2}}
\newcommand{\ithree}{{\it 3}}
\newcommand{\ifour}{{\it 4}}

\newcommand{\iten}{{\it 10}}

\newcommand{\fc}{\frac}
\newcommand{\ra}{\rightarrow}
\newcommand{\la}{\lambda}

\font\mybb=msbm12 at 12pt
\def\bb#1{\hbox{\mybb#1}}

\def\bE {\bb{E}}
\newcommand{\eei}{\mbox{$\bb{E}^8$}}

\title{Supergravity Supertubes}

\author{Roberto Emparan \thanks{Also at Departamento de F\'\i sica
Te\' orica, Universidad del Pa\'\i s Vasco, Bilbao, Spain.} \\
    Theory Division, CERN \\
    CH-1211 Geneva 23 \\
    Switzerland \\ 
    E-mail: \email{Roberto.Emparan@cern.ch}}

\author{David Mateos and Paul K.\ Townsend \\
   Department of Applied Mathematics and Theoretical Physics\\
   Centre for Mathematical Sciences \\
   Wilberforce Road, Cambridge CB3 0WA, United Kingdom \\
E-mail: \email{D.Mateos@damtp.cam.ac.uk, P.K.Townsend@damtp.cam.ac.uk}}

\abstract{
We find the supergravity solution sourced by a supertube: a
(1/4)-supersymmetric D0-charged IIA superstring that has been blown up
to a cylindrical D2-brane by angular momentum.
The supergravity solution captures all essential features of the
supertube, including the D2-dipole moment and an upper bound on the
angular
momentum: violation of this bound implies the existence of closed timelike
curves, with a consequent ghost-induced instability of supertube probes.}

\keywords{D-branes, Supersymmetry and Duality,
Brane Dynamics in Gauge Theories}

\preprint{CERN-TH/2001-159 \\ DAMTP-2001-49 \\ \tt{hep-th/0106012}}

\begin{document}
\section{Introduction}

It has been recently shown by two of us that a cylindrical D2-brane can be
supported against collapse by the angular momentum generated by electric
and magnetic Born-Infeld fields \cite{supertube}. These fields can be
interpreted as some number $q_s$ of dissolved IIA superstrings and some
number per unit  length $q_0$ of dissolved D0-branes. The angular
momentum $J$ is proportional to $q_0 q_s$ and the energy is minimized for
a cylinder radius proportional to $\sqrt{q_0q_s}$. Somewhat surprisingly,
this configuration preserves 1/4 of the supersymmetry of the IIA Minkowski
vacuum, hence the name {\it supertube}. As discussed in \cite{supertube},
and extended here to the case of $N>1$ D2-branes\footnote{Or to a single
$N$-wound D2-brane; for the purposes of this paper both situations are
equivalent.}, one can also consider a combined system consisting of a
supertube and D0-charged strings with given total string and D0
charges $Q_s$ and $Q_0$, respectively. In this case supersymmetry does not
fix the angular momentum but instead implies the upper bound
$|J| \le R |Q_s Q_0|^{1/2}$, with the radius of an $N$-times-wound
supertube given by $R^2 = |J| / N$.

The gravitational back-reaction was not considered in \cite{supertube},
but when this is taken into account a supertube becomes a source for
the IIA supergravity fields that govern the low-energy limit of the
closed string sector of IIA superstring theory.  Although this source is a
distributional one (in the limit of weak string coupling), one might
still expect it to generate a solution of IIA supergravity that is
non-singular everywhere away from the source.  Such a solution should
carry the same string and D0-brane charges as the D2-brane supertube,
as well as the same angular momentum, and
preserve the same 1/4 of the supersymmetry of the IIA vacuum. The aim of
this paper is to exhibit this solution, which we call the {\it
supergravity supertube}, and to study its properties.

As we shall see, the supergravity solution accurately reproduces the
features of the D2-brane supertube described above. In particular, the
bound on the angular momentum arises from the requirement of causality:
if $J$ exceeds the bound then the Killing vector field associated
with the angular momentum becomes timelike in a region near the supertube;
since its orbits are closed this implies the existence of closed timelike
curves (CTCs). This is a {\it global}  violation of causality that cannot
cause unphysical behaviour of any  {\it local} probe (as we verify for a
D0-brane probe) but it can and  does cause unphysical behaviour for
D2-brane probes, due to the appearance of a ghost-excitation on the
D2-brane worldvolume.

Since the D2-brane supertube is a supersymmetric solution one might expect
the force between parallel or concentric\footnote{There is no
topological distinction between these cases in a space of dimension
$\ge4$.} supertubes to vanish, allowing them to be superposed.
This is certainly the case for the D0-charged superstrings to which the
supertube reduces in the limit of zero angular momentum, and it has
been argued in the context of matrix theory \cite{koreans} that it is
also true of supertubes. As any force between parallel supertubes is
transmitted by the IIA supergravity fields, it would seem necessary to
consider the supergravity supertube solution to verify this, and this is
one motivation for the present work.  In fact, we shall exhibit
`multi-tube' solutions representing a number of parallel supertubes with
arbitrary locations and radii, which implies the existence of a
`no-force' condition between parallel supertubes. By considering D0 and
IIA string probes in this background we also establish a `no-force
condition' between supertubes and strings and D0-branes.

As noted in \cite{supertube}, the D2-brane supertube is T-dual to a
helical rotating IIB \mbox{D-string} (the S-dual of which is T-dual to a
helical rotating IIA string). These \mbox{(1/4)-supersymmetric} rotating
helical strings have since been studied in two recent papers
\cite{recent1,recent2}. A supergravity solution representing the
asymptotic fields of a IIB rotating helical D-string was also presented
in \cite{recent2}. In fact, this solution is T-dual to a six-dimensional
version of the supertube; we shall comment further on lower-dimensional
supertubes in a concluding section. We should also mention
that a IIA supergravity solution for a D2-brane tube with D0-branes
and IIA string charges was constructed previously by one of the
authors \cite{Emparan}; however, as already noted in \cite{Emparan}
this solution does not describe a supertube because it has no angular
momentum\footnote{The results we present here suggest that the solution
of \cite{Emparan} should be interpreted as a simple superposition
of an unstable D2-brane tube with D0-charged strings.}.

\section{Worldvolume Supertubes}
\label{review}

In this section we will review the results of \cite{supertube} that
are relevant to this paper, with a
slight extension to allow for multiply-wound D2-branes. The conventions
used here differ slightly from those of \cite{supertube}.

The starting point is a D2-brane in the ten-dimensional Minkowski
vacuum of IIA superstring theory. We write the Minkowski metric
as
\be
ds^2_\iten = -dt^2 + dx^2 + dr^2 + r^2 \, d\vp^2 + ds^2(\bE^6) \,,
\label{mink}
\ee
where $\vp \sim \vp + 2\pi$. The induced metric $g$ on a cylindrical
D2-brane  of constant radius $R$, at a fixed position in $\bE^6$, aligned
with the
$x$-direction and with cross-section parametrized by $\vp$, is
\be
ds^2(g) = -dt^2 + dx^2 + R^2 \, d\vp^2 \,,
\ee
where we have identified the worldvolume time with $t$. We will allow
for a time-independent electric field $E$ in the $x$-direction,  and a
time-independent magnetic field $B$, so the Born-Infeld 2-form is
\be
F = E \, dt \w dx + B \, dx \w d\vp \,.
\label{fs}
\ee

The number of supersymmetries preserved by any brane
configuration in a given spacetime is the number of independent
Killing spinors $\e$ of the background for which
\be
\Gamma \e = \e \,,
\label{kappa0}
\ee
where $\Gamma$ is the matrix appearing in the `$\kappa$-symmetry'
transformation of the worldvolume spinors, its particular form
depending on the background and on the type of brane. The spacetime
Minkowski metric \eqn{mink} may be written as
\be
ds^2_\iten = -e^t e^t + e^x e^x + e^r e^r + e^\vp e^\vp + e^a e^a
\ee
for orthonormal 1-forms
\be
e^t =dt \sac  e^x = dx \sac e^r = dr \sac
e^\vp = r\, d\vp \sac e^a =d\rho^a \,,
\ee
where $\{ \rho^a \}$ are Cartesian coordinates on $\bE^6$.
Let $\Gamma_t, \Gamma_x, \Gamma_r, \Gamma_\vp$ and $\{\Gamma_a\}$ be the
ten constant tangent-space Dirac matrices associated to the above
basis of 1-forms, and let $\Gamma_\natural$ be the constant matrix of
unit square which anticommutes with all them. The Killing spinors in
this basis take the form $\e = M_+ \, \e_0$, where $\e_0$ is a
constant 32-component spinor and
\be
M_\pm \equiv \exp\left( \pm {1\over2} \vp \, \Gamma_{r\vp} \right) \,.
\label{mpm}
\ee
For the D2-brane configuration of interest here we have
\be
\Gamma = \frac{1}{\sqrt{-\det(g+F)}} \,
\Big( \Gamma_{tx\varphi} + E \, \Gamma_{\varphi} \Gamma_\natural +
B \, \Gamma_{t}\Gamma_\natural \Big) \,,
\ee
where
\be 
\sqrt{-\det(g+F)} = \sqrt{1 - E^2 + B^2} \,.
\ee
The condition for supersymmetry \eqn{kappa0} thus becomes
\be
M_+ \, \left[ B \, \Gamma_t \Gamma_\natural
- \sqrt{-\det(g+F)} \right] \epsilon_0
+ M_- \, \Gamma_\vp \Gamma_\natural \,
\left[ \Gamma_{tx} \Gamma_\natural + E \right] \epsilon_0 =0 \,.
\label{susy0}
\ee
Since this equation must be satisfied for all values of $\vp$, both
terms must vanish independently. The vanishing of the second
term requires that $E=\pm 1$ and
$\Gamma_{tx} \Gamma_\natural \e_0 = \mp \e_0$. Without loss of
generality we choose $E=1$ and
\be
\Gamma_{tx} \Gamma_\natural \e_0 = - \e_0 \,.
\label{E}
\ee
Now the first term in \eqn{susy0} vanishes identically if $B=0$, in
which case we have 1/2 supersymmetry. We shall therefore assume that
$B \neq 0$. In this case vanishing of the first term requires
$\Gamma_{t} \Gamma_\natural \e_0 = \pm \e_0$ and  $\mbox{sign}(B) = \pm
1$. Again without loss of generality we shall assume that $B>0$ and
\be
\Gamma_{t} \Gamma_\natural \e_0 = \e_0 \,.
\label{B}
\ee
The two conditions \eqn{E} and \eqn{B} on $\e_0$ are compatible and
imply preservation of 1/4 supersymmetry. They are respectively
associated with string charge along the $x$-direction and with
D0-brane charge.

Under the conditions above, the D2-brane Lagrangian (for unit surface
tension) is
\be
\label{lag}
{\cal L} = -\sqrt{R^2 \, (1-E^2) + B^2} \,.
\ee
The momentum conjugate to $E$ takes the form
\be
\Pi \equiv \fc{\pa \call}{\pa E} =
\fc{R^2 \, E}{\sqrt{R^2 \, (1-E^2) + B^2}} \,,
\label{pi-flat}
\ee
and the corresponding Hamiltonian density is
\be
{\cal H} \equiv \Pi E - {\cal L} =
R^{-1} \sqrt{\left( \Pi^2 + R^2 \right) \left( B^2 + R^2 \right)} \,.
\label{H}
\ee
The integrals 
\be
q_s \equiv \fc{1}{2\pi} \oint d\varphi \, \Pi \qquad \mbox{and} \qquad
q_0 \equiv \fc{1}{2\pi} \oint d\varphi \, B
\label{charges}
\ee
are (for an appropriate choice of units) the IIA string conserved
charge and the \mbox{D0-brane} conserved charge per unit length carried by
the tube. For a supersymmetric configuration $E=1$ and $B$ is
constant, so from \eqn{pi-flat} and \eqn{charges} we deduce that
\be
R = \sqrt{|q_s q_0|} \,.
\label{radius}
\ee
The tension or energy per unit length of the tube is in turn
\be
\tau = \fc{1}{2\pi} \oint d\varphi \, {\cal H} \,.
\ee
This is of course minimized at the supersymmetric radius \eqn{radius},
for which we find
\be
\tau = |q_s| + |q_0| \,.
\label{tension}
\ee 
This result shows that the positive energies associated to D2-brane
tension and rotation are exactly cancelled by the negative binding energy
of the strings and D0-branes with the D2-brane, and hence that the
supertube is a genuine bound state. As we shall see later, this has
some counter-intuitive consequences. 

The crossed electric and magnetic fields generate a Poynting
2-vector-density with
\be
{\cal J}_\vp = \Pi B
\ee
as its only non-zero component. The integral of ${\cal J}_\varphi$
over $\varphi$ yields an angular momentum per unit length
\be
J=q_s q_0
\label{J}
\ee 
along the axis of the cylinder. It is this angular momentum that
supports the tube against collapse at the constant radius \eqn{radius}.
In ten dimensions, the angular momentum 2-form $L$ may have rank at
most 8. This rank is 2 for the supertube, $J$ being the only non-zero
skew-eigenvalue of $L$. Note also that the angular momentum selects a
2-plane in the 8-dimensional space transverse to the strings, where
the cross-section of the cylinder lies.

At this point we wish to consider a slight generalization of the results
of \cite{supertube} to allow for the possibilty of $N$ coincident
D2-brane tubes, or a single D2-brane supertube wound $N$ times
around the $\vp$-circle, or combinations of coincident and multiply-wound
D2-brane tubes. In any of these cases the local field theory on the
D2-branes will be a $U(N)$ gauge theory
\footnote{Provided that the
radius of the tube is much larger than the string scale. Around the
string scale a complex field coupling to the $U(1)$ factor arising
from strings connecting opposite points on the $\vp$ circle may
become tachyonic, in which case the tube would decay to the vacuum
with the emission of strings, D0-branes and massless closed string modes
carrying the angular momentum; this is possible because the circle
parameterized by $\vp$ is topologically trivial in spacetime.}. 
The results
obtained in this paper will depend only on the total number $N$ of
D2-branes, so we need not distinguish between these possibilities; for
convenience we shall consider the configuration of $N$ D2-branes as a
single $N$-times-wound D2-brane. In this case $\Pi$ and $B$ must be
promoted to matrices in the Lie algebra of $U(N)$, and their relationship
to the string and D0-brane charges becomes
\be
q_s \equiv \fc{1}{2\pi} \oint d\varphi \, \mbox{Tr} \, \Pi \sac
q_0 \equiv \fc{1}{2\pi} \oint d\varphi \, \mbox{Tr} \, B \,.
\label{charges-N}
\ee
Since only their components along the identity contribute, we shall
assume that only these components are non-zero and we shall still
denote them by $\Pi$ and $B$. We then have
\be
q_s = N \Pi \sac q_0 = N B \,.
\ee
In the case that a single D2-brane winds $N$ times around the $\vp$-circle
these relations are simply understood as due to the effective length
of the D2-brane now being $2\pi R N$.

The Hamiltonian density is now
\be
\calh = N \, R^{-1}
\sqrt{\left( \Pi^2 + R^2 \right) \left( B^2 + R^2 \right)} \,,
\ee
which is still minimized by $R=\sqrt{|\Pi B|}$. However, in terms of
the charges we now find
\be
R = \fc{\sqrt{|q_s q_0|}}{N} \,,
\label{radius-N}
\ee
and similarly for the angular momentum:
\be
J = \fc{|q_s q_0|}{N} \,.
\ee
Note however that the tension is still given by \eqn{tension}.

As discussed in \cite{supertube} for the $N=1$ case, the quantity $|Q_s
Q_0|/N$ is actually an {\it upper bound} on the angular momentum of a
supersymmetric state with given string and D0-brane charges
$Q_s$ and $Q_0$. For
\be
|J| \leq {|Q_s Q_0|\over N}
\label{bound}
\ee
there exist `mixed' configurations consisting of an $N$-times-wound
D2-brane supertube with charges $q_s$ and $q_0$ such that $J=q_s q_0/N$,
together with parallel IIA strings and D0-branes (and possibly D0-charged
IIA
strings) with charges $q_s'$ and $q_0'$ such that \mbox{$Q_s = q_s +
q_s'$} and
\mbox{$Q_s = q_0 + q_0'$}. Since the D0-branes and strings carry no
angular
momentum, the combined system possesses total angular momentum
$q_s q_0/N$, which is less than the maximal value allowed by the bound.
By transferring charge from the strings and D0-branes to the supertube one
can increase the angular momentum at fixed $Q_0$ and fixed $Q_s$, while
maintaining the 1/4 supersymmetry. However, this process can only be
continued until $q_s = Q_s$ and $q_0 = Q_0$. If one were to continue 
beyond this point, obtaining a configuration with $|q_s| > |Q_s|$, then
$q_s$ and $q_s'$ would have different signs, and their associated
supersymmetry projectors would have no common eigenspinors.

Thus we conclude that the angular momentum of a supersymmetric
state of IIA string theory with charges $Q_s$ and $Q_0$ is bounded from
above as in \eqn{bound}. Since the supersymmetric radius of the supertube
is always 
\be
R^2 = {|J|\over N} \, ,
\label{RJN}
\ee
the bound on the angular momentum may be rewritten as
\be
J^2 \leq R^2 \, |Q_sQ_0| \,.
\label{JRQ}
\ee

\section{Supergravity Supertubes}

Our starting point will be a solution of $D$=11 supergravity found in
\cite{GMT} that describes the intersection of two rotating M5-branes and
an M2-brane, with an M-wave along the string intersection. Although the
generic solution of this type preserves only 1/8 of the 32 supersymmetries
of the M-theory vacuum, the special case in which we set to zero the
M5-brane charges preserves 1/4 supersymmetry. The metric and 4-form field
strength of this solution are (in our conventions)
\bea
ds_{\it 11}^2 &=& U^{-2/3} \, \Big[ -dt^2 + dz^2 + K \, (dt + dz)^2 +
2 \, (dt + dz) \, A  + dx^2 \Big] +
U^{1/3} d\vec{y} \cdot d\vec{y} \,, \nn
F_{\it 4} &=& dt \wedge d U^{-1} \wedge dx \wedge dz \,- \,
(dt + dz) \wedge dx \wedge d \, (U^{-1}  A) \,.
\label{11d}
\eea
Here $\vec{y}=\{y^i\}$ are Cartesian coordinates on $\bE^8$. The membrane
extends along the $x$- and $z$-directions, and the wave propagates along
the $z$-direction. The functions $U$ and $K$ and the 1-form $A$ depend
only on $\vec{y}$. $U$ and $K$ are harmonic on $\bE^8$ and are
associated with the membrane and the wave, respectively.
$A$ determines the angular momentum of the solution, and its field
strength $dA$ satisfies the source-free Maxwell-like equation
\be
d *_8 dA = 0 \,,
\label{maxwell}
\ee
where $*_8$ is the Hodge dual operator on $\bE^8$.
The reduction to ten dimensions of this solution along
the $z$-direction yields a (1/4)-supersymmetric IIA supergravity solution:
\bea
ds^2_{\it 10} &=& - U^{-1} V^{-1/2} \, ( dt - A)^2 +
U^{-1} V^{1/2} \, dx^2 + V^{1/2} \, d\vec{y} \cdot d\vec{y} \,, \nn
B_{\it 2} &=& - U^{-1} \, (dt - A) \wedge dx + dt\wedge dx\,, \nn
C_{\it 1} &=& - V^{-1} \, (dt - A) + dt \,, \label{solution} \\
C_{\it 3} &=& - U^{-1} dt\wedge dx \wedge A \,, \nn
e^\phi &=& U^{-1/2} V^{3/4} \,, \nonumber
\eea
where $V=1+K$, and $B_{\it 2}$ and $C_{\it p}$ are the Neveu-Schwarz
and Ramond-Ramond potentials, respectively, with
gauge-invariant field strengths
\be
H_{\it 3} = d B_{\it 2} \sac F_{\it 2} = d C_{\it 1} \sac
G_{\it 4} = dC_{\it 3} - d B_{\it 2} \w C_\ione \,.
\ee
We have chosen a gauge for the potentials $B_2$, $C_1$ and $C_3$ such
that they all vanish at infinity if $U$ and $V$ are chosen such that
the metric at infinity is the $D$=10 Minkowski metric in canonical
Cartesian coordinates.  We note here for future use that
\be\label{G4}
G_{\it 4} =  U^{-1} V^{-1} \, (dt - A) \wedge dx \wedge dA \,.
\ee 
 
The Killing spinors of the solution \eqn{solution} follow from those
in eleven dimensions given in \cite{GMT}, but we have also verified them
directly in $D$=10 with the conventions of \cite{new}. They take a simple
form when the metric is written as
\be
ds_{\it 10}^2 = - e^t e^t + e^x e^x + e^i e^i \,,
\label{orthonormal}
\ee
with the orthonormal 1-forms
\be
e^t = U^{-1/2} V^{-1/4} \, ( dt - A) \sac e^x =  U^{-1/2} V^{1/4} \, dx
\sac e^i = V^{1/4} \, dy^i \,.
\ee
In this basis, the Killing spinors become
\be
\e = U^{-1/4} V^{-1/8} \, \e_0 \,,
\label{killing}
\ee
where $\e_0$ is a constant 32-component spinor subject to the
constraints \eqn{E} and \eqn{B}. As mentioned before, these two
constraints preserve 1/4 supersymmetry and are those associated
with a IIA string charge aligned with the $x$-axis and D0-brane
charge, as required for a supertube, although we still must specify
the functions $U$ and $V$ and the 1-form $A$ before we can make this
identification. 

To motivate the choice for these functions, consider first the
solution describing D0-charged fundamental strings located at a point
$\vec{y} = \vec{y}_\a$ in $\bE^8$ and aligned with the $x$-axis.
This solution is given by \eqn{solution} with
\be
U= 1 + {|Q_s| \over 6\Omega \, \left| \vec{y} -\vec{y}_\a \right|^6} \sac
V= 1 + {|Q_0| \over 6\Omega \, \left| \vec{y} -\vec{y}_\a \right|^6} \sac
A = 0 \,,
\label{choices0}
\ee
where $\Omega$ is the volume of the unit 7-sphere.
The constant $Q_s$ is the string
charge, while $Q_0$ is the D0-brane charge per unit length; this can be
seen
from their asymptotic contributions to their respective field strengths
$H_{\it 3}$ and $F_{\it 2}$. (The signs of these charges are flipped by
taking $t\rightarrow -t$ and/or $x\rightarrow -x$.)
\FIGURE{
\epsfig{file=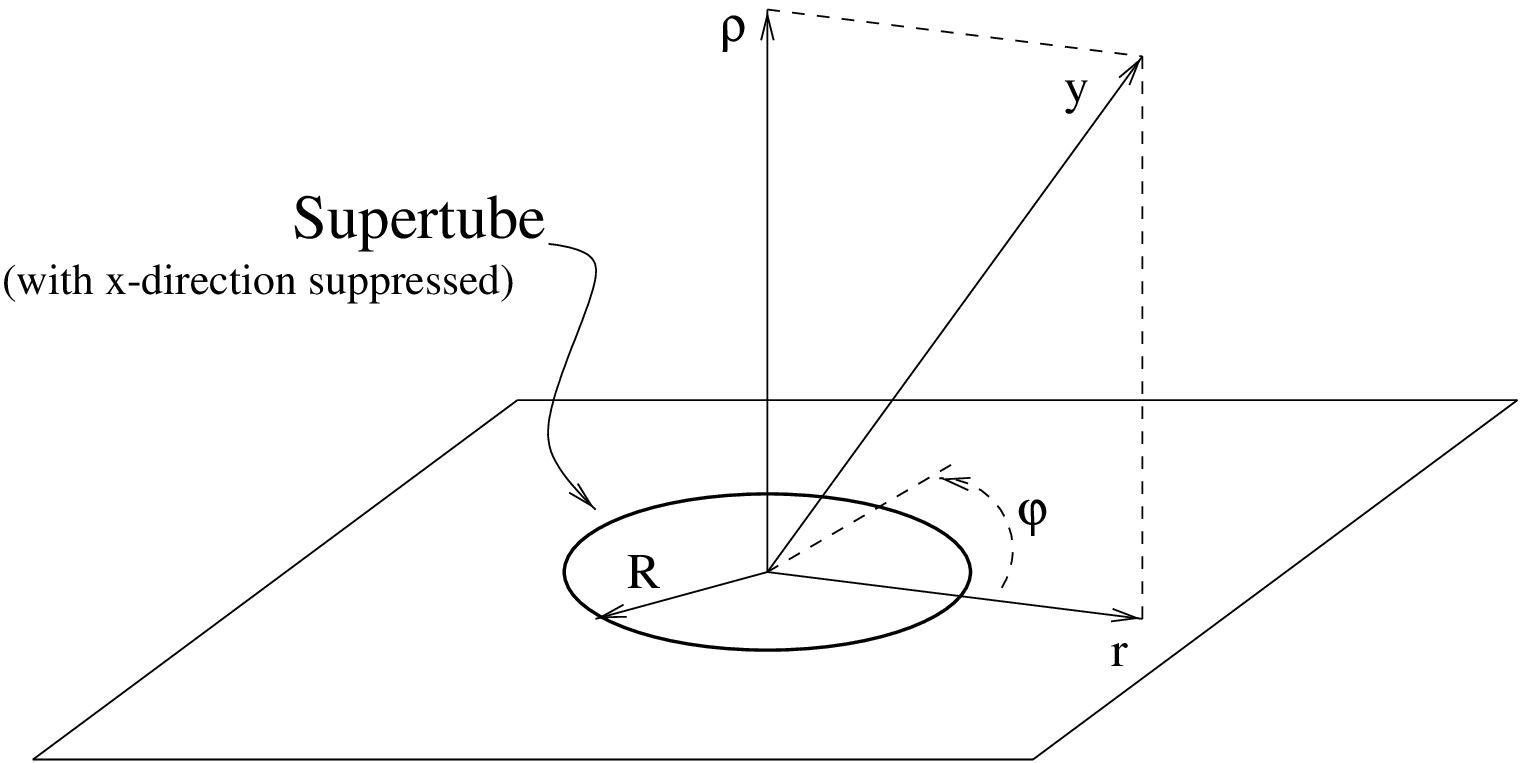, height=7cm}
\caption{Coordinates on \eei.}
\label{fig-coordinates}
}

Now consider distributing the charged strings homogeneously on a
circle of radius $R$ in a 2-plane in \eei. Let $r,\vp$ be polar
coordinates on this plane, and $\rho$ the radial coordinate on the
orthogonal 6-plane (see figure \ref{fig-coordinates}).
The metric then takes the form
\be
ds^2(\bE^8) = dr^2 + r^2d\varphi^2 + d\rho^2 +
\rho^2 d\Omega_{\it 5}^2  \,,
\ee
where $d\Omega_{\it 5}^2$ is the $SO(6)$-invariant metric on the unit
5-sphere. The solution for this configuration is still as in
\eqn{solution} with $A=0$,
but with the harmonic functions obtained by linear superposition on
the circle of those in \eqn{choices0}, that is,
\bea
U(r, \rho) &=& 1 + \fc{|Q_s|}{6 \Omega} \, \fc{1}{2\pi} \, \int_0^{2\pi}
d\a \,
\fc{1}{\left| \vec{y} -\vec{y}_\a \right|^6} \nn
&=& 1 + \fc{|Q_s|}{6 \Omega} \, \fc{1}{2\pi} \, \int_0^{2\pi} d\a \,
\fc{1}{\left(r^2 + \rho^2 + R^2 - 2R r \, \cos \a \right)^3} \,,
\eea 
and similarly for $V$ with $Q_s$ replaced by $Q_0$.
Thus
\bea
U &=& 1 + \fc{|Q_s|}{6 \Omega}  \,
\fc{\left(r^2 + \rho^2 + R^2 \right)^2 + 2 R^2 r^2}{\Sigma^5} \,, \nn
V &=& 1 + \fc{|Q_0|}{6 \Omega}  \,
\fc{\left(r^2 + \rho^2 + R^2 \right)^2
+ 2 R^2 r^2}{\Sigma^5} \,,
\label{choices}
\eea
where
\be
\Sigma (r,\rho) = \sqrt{\left(r^2 + \rho^2 + R^2 \right)^2 - 4 R^2 r^2}\,.
\label{defsigma}
\ee 
Note that (by construction) these functions satisfy Laplace's equation
on \eei\ with Dirac delta-like sources with support on the
$r=R, \rho =0$ circle. In addition, they have the same asymptotic
behaviour as those in \eqn{choices0} in the limit $\la \ra \infty$,
where $\la = |\vec{y}|$ is the radial coordinate in \eei.
Hence the constants $Q_s$ and $Q_0$ in \eqn{choices} are again
the string and D0 charges.

The solution \eqn{solution} with $A=0$ and $U$ and $V$ as in
\eqn{choices} displays a tubular structure but possesses no angular
momentum, and the field strength $G_\ifour$ sourced by D2-branes
vanishes. To describe the supertube we must incorporate the angular
momentum, with the source of rotation being located at the tube.
We know that the asymptotic form must be
\be
A \sim {1\over2}{L_{ij} \, y^j \over \Omega \, \lambda^8} \, dy^i \,,
\ee
where the constants $L_{ij}=-L_{ji}$ are the components of the angular
momentum 2-form $L$. As explained above,
$L$ must have rank 2 for the supertube, in which case we may write the
asymptotic form as
\be
A\sim{J\over 2\Omega\lambda^8}r^2d\varphi \,,
\label{asympA}
\ee
where $J$ is a constant angular momentum, the one non-zero
skew-eigenvalue of $L$. The calculation for the exact form of $A$
sourced by the supertube is essentially the same as that giving the
vector potential created by a circular electric current of intensity
proportional to $J/R^2$, that is,
\be
A = {1\over 6\Omega}\, {J\over 2\pi R} \, r d\varphi \,
\int_0^{2\pi} d\a \, \fc{\cos \a}
{\left(r^2 + \rho^2 + R^2 - 2R r \, \cos \a \right)^3} \,.
\ee
The result is 
\be
A = J \, \fc{\left(r^2 + \rho^2 + R^2 \right) \, r^2}
{2 \Omega \, \Sigma^5}  \, d\vp  \,,
\label{A}
\ee
which has the correct asymptotic behaviour \eqn{asympA}. As we shall see
in the next section, this choice of $A$ automatically generates the
correct D2-fields. 

The supertube solution is thus given by \eqn{solution} with $U$ and
$V$ as in \eqn{choices} and $A$ as in \eqn{A}; we shall analyze its
properties in detail below. However, we wish to note here that
evaluation of the ADM integral for brane tension \cite{Stelle}
(in the same conventions as for the angular momentum) yields the
tube tension 
\be
\tau = |Q_s| + |Q_0| \,,
\label{tension1}
\ee
exactly as for the D2-brane supertube (no factors of the string coupling
constant appear in this formula because the asymptotic value of the
dilaton vanishes for our chosen solution).

Having completed our construction of the supertube solution,
we close this section by presenting its generalization to a
`multi-tube' solution representing $N$ parallel tubes with
arbitrary locations, radii and charges. These are easily constructed
because of the linearity of the harmonicity conditions on $U$ and $V$
and of the Maxwell equation for $A$. The general expression is
\bea
U &=& 1 + \sum_{n=1}^N
\frac{Q_s^{(n)}}{6 \Omega}\,
\fc{\left( |\vec{y} - \vec{y}_n|^2 + R_n^2 \right)^2 +
2 R_n^2 |\vec{r} - \vec{r}_n|^2}
{\Big[\left( |\vec{y} - \vec{y}_n|^2 + R_n^2 \right)^2 -
4 R_n^2 |\vec{r} - \vec{r}_n|^2 \Big]^{5/2}} \,, \nn
V &=& 1 + \sum_{n=1}^N
\frac{Q_0^{(n)}}{6 \Omega}\,
\fc{\left( |\vec{y} - \vec{y}_n|^2 + R_n^2 \right)^2 +
2 R_n^2 |\vec{r} - \vec{r}_n|^2}
{\Big[\left( |\vec{y} - \vec{y}_n|^2 + R_n^2 \right)^2 -
4 R_n^2 |\vec{r} - \vec{r}_n|^2 \Big]^{5/2}} \,, \\
A &=& \sum_{n=1}^N {J^{(n)} \over 2 \Omega} \,
\frac{\left( |\vec{y} - \vec{y}_n|^2 + R_n^2 \right)^2}
{\Big[\left( |\vec{y} - \vec{y}_n|^2 + R_n^2 \right)^2 -
4 R_n^2 |\vec{r} - \vec{r}_n|^2 \Big]^{5/2}}
\left[ (u-u_n) dv - (v-v_n) du \right] \,. \nonumber
\eea
Here $u$ and $v$ are Cartesian coordinates on the 2-plane selected by
the angular momentum. All the tubes are aligned along the
$x$-direction and all their cross-sections are parallel to each other.
The $n$-th tube has radius $R_n$ and is centred at
\be
\vec{y} = (\vec{r}_n, \vec{\rho}_n) = (u_n, v_n, \vec{\rho}_n)
\ee
in \eei. It carries string and D0 charges $Q_s^{(n)}$ and
$Q_0^{(n)}$, respectively, and angular momentum $J^{(n)}$.
The total charges and angular momentum are the sums of those carried
by each tube. Clearly, by setting some of the radii and angular
momenta to zero, we obtain a solution representing a
superposition of D0-charged strings and supertubes.

The existence of this multitube solution shows that there is no force
between stationary parallel supertubes
\footnote{Nor between supertubes and strings or D0-branes. We shall
confirm and elaborate on this in section \ref{susy-sec}.}.
Note that, as far as the
supergravity solution is concerned, two (or more) tubes can intersect
each other, as described by the solution above when the radii and centres
are chosen appropriately.

\section{D2-Dipole Structure and Closed Timelike Curves}
\label{structure}

We have now found a (1/4)-supersymmetric solution of IIA supergravity that
carries all the charges required for its interpretation as the solution
sourced by a supertube source. Although it also displays the
appropriate tubular structure, we have not yet identified clearly the
presence of a D2-brane. Even though there cannot be any D2-brane charge,
we would still expect the fields of the D2-brane supertube to carry a
non-zero D2-brane dipole moment determined by the size of the tube. There
is
certainly a local D2-brane charge distribution because the electric
components of $G_{\it 4}$ are non-zero. Specifically, for $A$ given by
(\ref{A}), we have, asymptotically,
\be
G_{\it 4} = d \, \left[
\frac{Jr^2}{2\Omega (r^2+\rho^2)^4} \,
dt \w dx \w d\varphi \right] + \ldots \,,
\label{dipole-field}
\ee
where the dots stand for subleading terms in an expansion in
$1/\lambda$ for $\lambda \rightarrow \infty$. The integral of
$*G_{\it 4}$ over any 6-sphere at infinity vanishes, showing that
there is no D2-brane net charge. However, the expression
\eqn{dipole-field} has the correct form to be interpreted
as the dipole field sourced by a cylindrical D2-brane aligned with
the $x$-axis; the scale of the dipole moment is set by the
angular momentum,
\be
\mu_{\it 2} \sim J \,.
\label{dipole-moment}
\ee
In turn, the dipole moment must be related to the size of the source as
\be
|\mu_{\it 2}| \sim N R^2 \,,
\label{radius1}
\ee
where $R$ is the radius of the D2-tube and $N$ the number of
D2-branes. This can be seen as follows. The dipole moment of $N$
spherical D2-branes of radius $R$ scales as $|\mu| \sim N R^3$
\cite{Myers}. For a cylindrical D2-brane of length $a$ this must be
replaced by $|\mu| \sim N a R^2$. In the present
situation $a$ is infinite, so $\mu_{\it 2}$ in equations
\eqn{dipole-moment} and \eqn{radius1} is actually a dipole moment
per unit length. Thus, we conclude that
\be
R^2 \sim \fc{|J|}{N} \,,
\label{radius2}
\ee
in agreement with the worldvolume analysis of section \ref{review}.
This agreement could be made more precise by fixing the proportionality
factors in the relations above. Instead, we now turn to analyzing
the solution near the tube, which will yield the same result and
provide additional information on the detailed structure of the
supertube, in particular concerning the presence and the number of
D2-branes. 

To this effect, we perform a change of coordinates which is 
convenient to focus on the region close to the tube:
\bea
r&=&\sqrt{(\hat{r}\cos\hat\theta+R)^2-\hat{r}^2} \,, \nn
\rho&=&\hat{r}\sin\hat\theta \,.
\eea
This has been designed so that, in the new coordinates,
$\Sigma=2R\hat{r}$. In the limit
\be
\hat{r}/R \ll 1
\label{hat-limit}
\ee
one approaches the tube at $r=R$, $\rho=0$. Note that this can be
achieved by either fixing $R$ and making $\hat{r}$ small, or by fixing
$\hat{r}$ and making $R$ large. In the first case one focuses, for a
given solution, on the region near the tube. In the second, the radius
of the tube grows very large while we remain at a finite distance from
it. In both cases the tube looks planar. In this limit the metric becomes
\bea
ds^2 &=& -U^{-1}V^{-1/2}\, (dt- k \, d \hat{z})^2
+ U^{-1}V^{1/2} \, dx^2 \nn
&& + V^{1/2} \, \left( d\hat{z}^2+d\hat{r}^2+\hat{r}^2 d\hat\theta^2+
+\hat{r}^2\sin^2\hat\theta d\Omega_{\it 5}^2\right)\,,
\label{close-up-solution}
\eea
where we have defined $\hat{z}=R\varphi$, so $\hat{z}$ is a coordinate
identified with period $2\pi R$. The three functions in the metric
above are 
\bea 
U &=& 1+ {|Q_s|/2\pi R\over 5\Omega_{\it 6} \hat{r}^5} + \cdots \,,\nn
V &=& 1+ {|Q_0|/2\pi R\over 5\Omega_{\it 6} \hat{r}^5} + \cdots \,, \\
k &=& {J/2\pi R^2\over 5\Omega_{\it 6} \hat{r}^5} + \cdots \,, \nonumber
\label{close-solution}
\eea
where the dots stand for subleading $\hat{r}$-dependent terms in the
limit \eqn{hat-limit}. Here $\Omega_{\it 6}$ is the volume of the unit
6-sphere. The gauge potentials are as in \eqn{solution} with $U$
and $V$ given above and $A=k d\hat{z}$.

The solution in the form \eqn{close-up-solution} clearly exhibits
the properties expected from the planar limit close to the supertube.
The angular momentum becomes linear momentum along the tangent direction
to
the circle, that is, along $\hat{z}$. The $SO(6)$-symmetry associated 
to rotations in $\bE^6$ is enhanced to $SO(7)$: the
$\hat{\theta}$-coordinate in the near-tube metric combines with the 
coordinates on the 5-sphere to yield the metric on a round 6-sphere.
The functions $U$ and $V$ are sourced by delta-functions at
$\hat{r}=0$, hence the gauge potentials $B_{\it 2}$ and
$C_{\it 1}$ correspond to charge densities $Q_s/2\pi R$ and
$Q_0/2\pi R$ along the $\hat{z}$-direction.
The four-form field strength
\be
G_{\it 4}=-U^{-1}V^{-1}\,dt\wedge dx\wedge d\hat{z}\wedge dk
\ee 
corresponds to the charge density of $N=|J|/R^2$ D2-branes at
$\hat{r}=0$, in agreement with \eqn{RJN}.
To see this, consider a 7-disc that is small enough to
intersect the \mbox{D2-circle} at only one
point (see figure \ref{6-sphere}) and compute the flux of $*G_{\it 4}$
through its 6-sphere boundary. The result is precisely
\be
N \equiv \left| 2\pi \int_{S^6}*G_{\it 4}\right| = {|J| \over R^2} \,.
\ee
As the radius of the 6-sphere increases, the 7-disc will
eventually intersect the D2-circle at a second point and the above integral
then vanishes, as happens for any 6-sphere at infinity
\footnote{It follows from the considerations above that 
the limit $R\to\infty$ with 
fixed charge densities results in a planar configuration of D2-branes with
strings and D0-branes distributed on it, and with momentum in the
direction transverse to the strings. This preserves the same amount of
supersymmetries as the supertube. To our knowledge,
this configuration has not been previously considered.}.
\FIGURE{
\epsfig{file=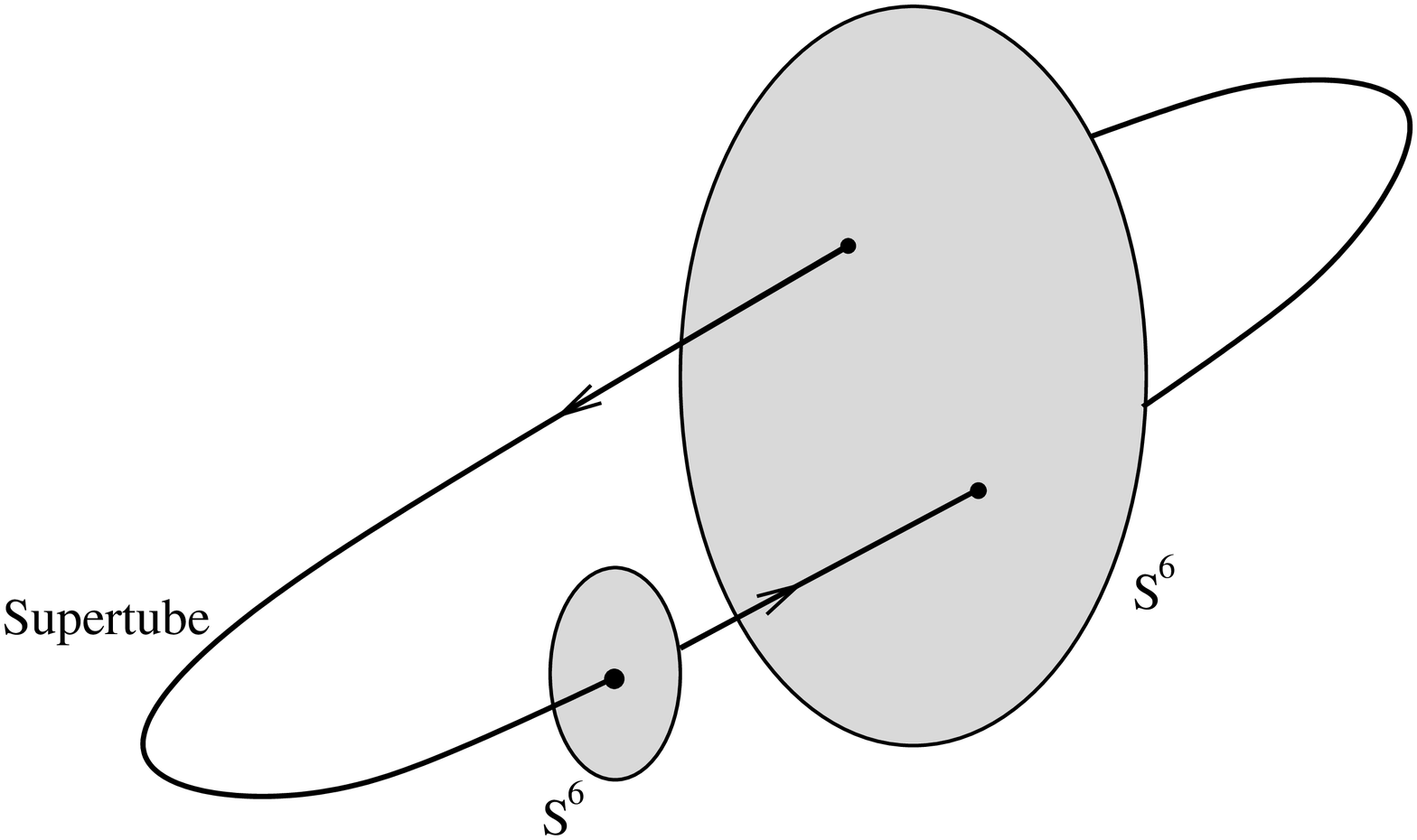, height=7cm} 
\caption{The little 7-disc intersects the supertube (suppressing the 
$x$-direction) at one point, hence the flux of $*G_{\it 4}$ through its
boundary (the little $S^6$) is non-zero. On the contrary, the large
disc intersects the supertube at two points which correspond to local
D2-brane charge densities of opposite sign, hence the flux through 
the large $S^6$ vanishes; the same happens for any 6-sphere at infinity.}
\label{6-sphere}  
}

We have now identified all the elements present in the supertube solution.
There are four independent parameters, say $Q_s$, $Q_0$, $R$
and $N$ (or $J$). So far, the solution displays all the features of the
worldvolume supertube except for the bound \eqn{JRQ}. We shall now
show that this arises by demanding that there be no causality
violations in the supertube spacetime.

Consider the Killing vector field $\ell \equiv \pa / \pa \vp$, which is
associated to rotations on the 2-plane selected by the angular
momentum. Its norm squared is
\be
|\ell|^2 = g_{\varphi\varphi} = U^{-1}V^{-1/2}r^2 \,
\left(UV - f^2 / r^2 \right) \,,
\label{l2}
\ee
where 
\be
\label{eff}
f= J \, \fc{\left(r^2 + \rho^2 + R^2 \right) \, r^2}
{2 \Omega \, \Sigma^5}  \,.
\ee
The norm of $\ell$ is always positive for large enough $\la$,
regardless of the value of $J$, and if the bound \eqn{JRQ}
is satisfied then $\ell$ remains spacelike everywhere.
However, if $J$ exceeds the bound then $\ell$ becomes timelike
sufficiently close to the tube. To see this, we write out \eqn{l2} as
\bea
|\ell|^2 &=& {U^{-1} V^{-1/2} r^2 \over36 \Omega^2 \Sigma^{10}}
\Biggl\{ 36\Omega^2 \Sigma^{10} 
+ 6\Omega \Sigma^5 (|Q_0|+|Q_s|) \left[(r^2+\rho^2+R^2)^2 + 2R^2r^2\right]
\nn
&+& |Q_0Q_s| \Biggl[ (r^2-R^2)^2(r^4+r^2R^2+R^4)+2\rho^2(r^2+R^2)
\left[2(r^2-R^2)^2+3r^2R^2\right]\nonumber\\
&+&\rho^4\left[6(r^2-R^2)^2 +19
R^2r^2\right]+4\rho^6(r^2+R^2)+\rho^8\Biggr]\nn
&+&9(R^2 |Q_sQ_0| - J^2) (r^2+R^2+\rho^2)^2 r^2 \Biggr\} \,.
\eea
All the terms in this expression 
are non-negative except for the last
one, which becomes negative precisely when the bound \eqn{JRQ} is
violated. In addition, in the near-tube limit
\be
r \ra R \sac \rho \ra 0 
\label{limit}
\ee 
we find
\be
|\ell|^2 = \fc{R^2}{\Omega^{1/2} \Sigma^{5/2} |Q_s||Q_0|^{1/2}} \,
\Big[ \left( R^2 |Q_s Q_0| - J^2 \right) + \ldots \Big] \,,
\ee
where the dots represent terms which vanish in the limit \eqn{limit}. 
Thus the supergravity supertube has CTCs if and only if it is
`over-rotating'. These CTCs are `naked' in the sense that
there is no event horizon to prevent them from being deformed to pass
through any point of the spacetime. Neither can they be removed by a
change of frame, since the causal structure is unchanged by any
non-singular conformal rescaling of the metric. It should be appreciated
that the surface defined by $|\ell|^2=0$ (present when $J$ exceeds the
bound) is {\it not} singular; it is not even a coordinate singularity.
In particular, the metric signature does {\it not} change on this
surface despite the fact that for $|\ell|^2<0$ there are two commuting
timelike Killing vector fields. In fact, the only physical singularity
of the solution \eqn{solution} is at $r=R, \rho =0$, where both
$|\ell|^2$ and the dilaton diverge. Thus, the over-rotating
supertube spacetime is {\it locally} as physical as the under-rotating
one, with a metric that is singular only at the location of the tube.  

The CTCs of the over-rotating supertube metric have a very simple
origin in eleven dimensions. The IIA supergravity supertube lifts to the
solution \eqn{11d} of eleven-dimensional supergravity from which we
started, with $U$ and $V$ as in \eqn{choices} and $A$ as in \eqn{A}. If
the eleventh coordinate $z$ is not periodically identified then this
eleven-dimensional spacetime has no CTCs, but CTCs are created by the
identifications needed for compactification on a circle if $J$ exceeds
the bound \eqn{JRQ}. To see
this, consider an orbit of the vector field $\xi = \pa_\varphi + \beta
\pa_z$, for real constant $\beta$. When $z$ is periodically
identified this curve will be closed for some dense set of values of
$\beta$. It will also be timelike if $|\xi|^2<0$, which will happen if
\be
\beta^2 + 2V^{-2}f\, \beta + UV^{-2}r^2 < 0 \,.
\ee
This inequality can be satisfied for real $\beta$ only if
$r^2<U^{-1}V^{-1}f^2$, but this is precisely the condition that $\ell$ be
timelike. Thus, there will be closed timelike orbits of $\xi$ in the
$D$=11 spacetime precisely when the closed orbits of $\ell$ become
timelike
in the $D$=10 spacetime. However, from the eleven-dimensional perspective,
these CTCs arise from periodic identification and can therefore be
removed by passing to the universal covering space. An analogous
phenomenon was described in \cite{carlos}, where it was shown that the
CTCs of over-rotating supersymmetric five-dimensional black holes are
removed by lifting  the solution to ten dimensions and passing to the
universal covering space.

\section{Brane Probes: Supersymmetry}
\label{susy-sec}

We shall now examine the behaviour of brane probes in the supertube
spacetime. Specifically, we shall consider D0-branes, strings and
D2-branes. We shall first show that the 1/4 supersymmetry of the
background is preserved by stationary D0-brane, IIA-string and
D2-supertube probes (when suitably aligned).

The $D$=10 spacetime metric \eqn{solution} can be written as
\be
ds_{\it 10}^2 = - e^t e^t + e^x e^x + e^r e^r +
e^\vp e^\vp + e^a e^a \,,
\ee
for orthonormal 1-forms
\bea
e^t &=& U^{-1/2} V^{-1/4} \, ( dt - A) \,,\nn
e^x &=&  U^{-1/2} V^{1/4} \, dx \,, \nn
e^r &=& V^{1/4} \, dr \,, \\
e^\vp &=& V^{1/4} \, r \, d\vp \,, \nn
e^a &=& V^{1/4} \, d\rho^a \,. \nonumber
\label{ortho}
\eea
Here $\{\rho^a\}$ are Cartesian coordinates on the $\bE^6$ space
orthogonal to the 2-plane selected by the angular momentum.
The Killing spinors in this basis take the form
\be
\e = U^{-1/4} V^{-1/8} \, M_+ \, \e_0 \,,
\ee
where $\e_0$ is a constant 32-component spinor subject to the
constraints \eqn{E} and \eqn{B}. Recall that we introduced the
matrices  $\Gamma_t, \Gamma_x, \Gamma_r, \Gamma_\vp, \{\Gamma_a\},
\Gamma_\natural$ and $M_\pm$ in section \ref{review}. Recall too that
the number of supersymmetries preserved by any brane configuration in a
given spacetime is the number of independent Killing spinors of the
background verifying \eqn{kappa0}.

For a stationary D0-brane in the gauge in which worldline time
is identified with $t$ we have
\be
\Gamma_{D0} =  \Gamma_t \Gamma_\natural \,.
\ee
For a IIA string in the same temporal gauge and, additionally, with the
string coordinate identified with $x$, we have
\be
\Gamma_{string} = -\Gamma_{tx} \Gamma_\natural \,.
\ee
These two $\Gamma$-matrices commute and the Killing spinors of the
background are simultaneous eigenstates of $\Gamma_{D0}$ and
$\Gamma_{string}$ with unit eigenvalue, so the inclusion of these
D0-brane and IIA-string probes does not break any supersymmetries that
are preserved by the background. In particular, this means that a
D0-brane or a string can be placed arbitrarily close to the supertube
without experiencing any force. This result is counter-intuitive
because one would expect the D2-cylinder in this situation
to behave approximately like a planar D2-brane, and hence to exert an
attractive force on \mbox{D0-branes} and strings. 
The explanation is presumably
that the supertube is in fact a true bound state of strings,
D0-branes and cylindrical D2-branes, in which the D2-branes behave
genuinely differently as compared to their free counterparts\footnote{The
neutron considered as a bound state of quarks is analogous since the
attractive force between neutrons is much weaker than the force between
its constitutive quarks.}.

Now we consider a probe consisting of a supertube itself,
that is, a D2-brane of cylindrical topology with string and D0-brane
charges. The analysis is very similar to that of section \ref{review}
except that the Born-Infeld field strength $F$ must now be replaced
by the background-covariant field strength ${\cal F}=F - B_\itwo$.
As before we choose $F$ to have the form
\be
F= E \, dt\wedge dx + B \, dx\wedge d\varphi\, .
\ee
This yields
\be
{\cal F} = {\cal E} \, dt\wedge dx + {\cal B} \, dx\wedge d\varphi\, ,
\ee
where
\be
{\cal E} = E + U^{-1} -1 \sac {\cal B} = B + U^{-1}f\, .
\ee
In the orthonormal basis \eqn{ortho} we have
\be
\calf = \bcale \, e^t \w e^x + \bcalb \, e^x \w e^\vp \,,
\ee
where
\be
\bcale = U \cale \sac \bcalb = U^{1/2} V^{-1/2} r^{-1} \,
(\calb - f \cale) \,.
\ee
In terms of these variables, $\Gamma_{D2}$ takes the same form as in
flat space:
\be
\Gamma_{D2} = \frac{1}{\sqrt{-\det(g+\calf)}} \,
\Big( \Gamma_{tx\varphi} + \bcale \, \Gamma_{\varphi} \Gamma_\natural +
\bcalb \, \Gamma_{t}\Gamma_\natural \Big) \,,
\ee
where
\be 
\sqrt{-\det(g+\calf)} = \sqrt{1 - \bcale^2 + \bcalb^2} \,.
\ee
The condition for supersymmetry \eqn{kappa0} thus becomes
\be
M_+ \, \left[ \bcalb \, \Gamma_t \Gamma_\natural
- \sqrt{-\det(g+\calf)} \right] \epsilon_0
+ M_- \, \Gamma_\vp \Gamma_\natural \,
\left[ \Gamma_{tx} \Gamma_\natural + \bcale \right] \epsilon_0 =0 \,.
\label{susy}
\ee
Both terms must vanish independently. Given the constraints
\eqn{E} and \eqn{B} satisfied by $\e_0$, the vanishing of the second
term requires that $\bar{\cal E}=1$. Equivalently,
\be
\label{susycon} 
{\cal E} = U^{-1}\, .
\ee
The first term in (\ref{susy}) then vanishes identically if
$\bcalb \ge 0$, which using \eqn{susycon} is equivalent to $B\ge 0$.
In the $B=0$ case we have 1/2 supersymmetry, so we shall assume that
\be
B > 0 \,.
\label{susycon2}
\ee

\section{Brane Probes: Energetics}
\label{enrgetics}

We now turn to the energetics of probes in the supertube spacetime.
Our aim is to uncover any effects of the presence of CTCs on
brane probes, so in this section we shall assume that the bound
\eqn{JRQ} is violated. We shall see that the {\it global} causality
violation due to the CTCs causes a {\it local} instability on the
worldvolume of extended probes such as a D2-brane tube, the reason for
this being of course that the probe itself is non-local.
On the contrary, we would not expect any unphysical
effect on a local probe such as a D0-brane, and we shall
begin by verifying this.

The action for a D0-brane of unit mass is
\be
S_{D0} = - \int e^{-\phi} \sqrt{-\det g} - \int C_\ione \,.
\ee
In the gauge where the worldline time is identified with $t$ we find
the Lagrangian density
\be
\call = - V^{-1} \sqrt{ 1 - 2f \, \dot{\vp} -
U V^{1/2} \, (g_{\vp\vp} \, \dot{\vp}^2 + v^2)} +
V^{-1} (1 - f \dot{\vp} ) -1 \,,
\label{density}
\ee
where the overdot indicates differentiation with respect to $t$, $f$
is given by (\ref{eff}), and
\be
v^2 \equiv g_{ij} \dot{X}^i \dot{X}^j
\sac X^i = \{ x, r, \rho^a\} \,.
\ee
Note that $v^2 \geq 0$.

The Lagrangian \eqn{density} appears to lead to unphysical behaviour
if $g_{\vp\vp}$ becomes negative. However, an expansion in powers
of the velocities yields
\be
\call = -1 + \frac{1}{2} U r^2 \, \dot{\vp}^2 +
\frac{1}{2} U V^{-1/2} v^2 + \cdots \,.
\ee
The first term is minus the (unit) positive mass of the D0-brane. As
expected from the supersymmetry of a stationary D0-brane, all the
velocity-independent potential terms cancel. In addition, the kinetic
energy is positive-definite regardless of the sign of $g_{\vp\vp}$ (and
this remains true to all orders in the velocity). There is therefore
nothing to prevent a D0-brane from entering the region where
$|\ell|^2 < 0$, and its dynamics in this region is perfectly physical,
at least locally. The same
applies to $x$-aligned IIA superstrings; again this is not surprising because
this is a probe that is local in $\varphi$.

We now turn to the supertube itself. This is a more significant test of
the geometry because one could imagine building up the source of the
supergravity solution by accretion of concentric D2-brane supertubes.
We begin by considering a general tubular D2-brane aligned with the
background. For our purposes we may assume that
the worldvolume scalar fields $r$ and $\rho^a$ are uniform in $x$, but
we shall also assume, initially, that they are time-independent (as
will be the case for a supertube,
since this is a D2-brane tube at a minimum of the potential energy).

The action 
\be
S_{D2} = - \int e^{-\phi} \sqrt{-\det(g+\calf)} -
\int \left( C_\ithree + C_\ione \w \calf \right)
\ee
in the physical gauge (as used above) yields the Lagrangian density
\be
\label{lagphys}
{\cal L} = - U^{1/2}V^{-1} \sqrt{ V r^2 (U^{-2}-{\cal E}^2) +
U^{-1}({\cal B}-f{\cal E})^2} + V^{-1}({\cal B}-f{\cal E}) - B \,.
\ee
The variable conjugate to $E$ is
\be
\label{conjugate}
\Pi \equiv {\pa {\cal L}\over \pa E} = {\pa {\cal L}\over \pa
{\cal E}} = { U^{1/2} r^2 {\cal E} + U^{-{1/2}} V^{-1}
({\cal B}- f {\cal E}) f \over \sqrt{V r^2(U^{-2} - {\cal E}^2) +
U^{-1}({\cal B}-f{\cal E})^2}} - V^{-1} f \,.
\ee
The Hamiltonian density is defined as
\be
{\cal H} \equiv \Pi E -{\cal L} = \Pi{\cal E} - {\cal L} +
\Pi(1-U^{-1})\, .
\ee
For supersymmetric configurations we have ${\cal L}=-B$
and ${\cal E}= U^{-1}$, so
\footnote{Note that both $B$ (see equation \eqn{susycon2}) and $\Pi$
(from its definition \eqn{conjugate}) are positive for supersymmetric
configurations.}
\be
{\cal H}= \Pi + B
\ee
for a supertube. Neither $\Pi$ nor $B$ is invariant under background
gauge transformations. However, their worldspace integrals are
gauge-invariant
\footnote{We should consider only time-independent gauge
transformations because the Hamiltonian is not expected to be
invariant under time-dependent gauge transformations. We should also
consider only $x$-independent gauge transformations to be consistent
with our assumption of $x$-independence.}.
Given the assumption of \mbox{$x$-independence}, we can identify the IIA
string charge $q_s$ and the D0-brane charge per unit length $q_0$
carried by the supertube probe with the integrals of $\Pi$ and $B$
over $\vp$, as in \eqn{charges}. For constant $\Pi$ and $B$ we
therefore have
\be
\label{tubecharges}
\Pi = q_s\, ,\qquad B= q_0\, .
\ee
The probe supertube tension is then seen to be
\be\label{probeten}
\tau_{probe} = q_s+ q_0 \,.
\ee
Setting $\cale = U^{-1}$ in (\ref{conjugate}) we deduce that
\be
\Pi B = r^2
\ee
at the supertube radius, and hence that
\be
r = \sqrt{q_sq_0} \,.
\ee
The tension and the radius of a supertube probe in the supertube
spacetime are therefore exactly as in a Minkowski background (with
$N=1$). Note that this result holds regardless of whether or not CTCs
are present.

Thus, supersymmetry places no restriction on the possible radius of a
probe supertube in a supergravity supertube background. However,
the supertube minimizes the energy for given $q_s$ and $q_0$
(and hence is stable) only if there are no ghost excitations of the 
D2-brane, that is, excitations corresponding in
the quantum theory to negative norm states. These states {\it are} present 
for some choice of $q_s$ and $q_0$ whenever $g_{\varphi\varphi}<0$.
To establish this, we now allow for time-dependent $r$ and $\rho^a$,
and we expand the Lagrangian density in powers of
\be
v^2 = \dot{r}^2 + \delta_{ab} \, \dot{\rho}^a \dot{\rho}^b \,.
\ee
We find
\footnote{The calculation here is similar to the one performed in
\cite{JPP} but the interpretation is different.}
\be
{\cal L} = {\cal L}_0 + {1\over 2} M v ^2 + \cdots \,,
\ee
where ${\cal L}_0$ is the Lagrangian density of (\ref{lagphys}), and
\be
M= \fc{B^2 + 2U^{-1} f B + U^{-1}V r^2}
{U^{-1/2} V^{1/4} \, \sqrt{-\det (g+ {\cal F})}}  \,,
\ee
which reduces to
\be
M= \fc{B^2 +2U^{-1}f B + U^{-1}V r^2}{U^{-1}V^{1/2} B}
\ee
for a supertube. In both of these expressions for $M$, the denominator
is positive, and so is the numerator as long as
$g_{\varphi\varphi}>0$, but the factor
\be
B^2 + 2U^{-1}f B + U^{-1}V r^2
\ee
becomes negative for some choices of $B$ and $r$ (and hence $q_0$ and
$q_s$ for a supertube) whenever $g_{\varphi\varphi}<0$.

Note that the supertube does {\it not} become tachyonic;
its tension continues to be given by (\ref{probeten}). A tachyon
instability leads to runaway behaviour in which the kinetic energy
increases at the expense of potential energy. The instability here is
instead due to the possibility of negative kinetic energy, characteristic
of a ghost
\footnote{As representations of the Poincar{\'e} group, a `tachyon'
is a particle with spacelike energy-momentum vector, and hence negative
mass-{\it squared} (corresponding in the quantum theory to an
excitation about a vacuum that is a local maximum of the energy rather
than a local minimum).
A `ghost' is a particle with non-spacelike energy-momentum
vector but negative energy; at the level of the particle Lagrangian
(or, more generally, brane Lagrangian, as discussed here) this corresponds
to negative {\it mass}.}. This instability indicates that it is not
physically possible to construct a supertube spacetime with naked
CTCs starting from Minkowski space. One might imagine assembling such
a spacetime by accretion from infinity of `supertube shells' 
carrying infinitesimal fractions of charges
and angular momentum. When a finite macroscopic fraction of charges
and angular momentum have been accumulated, it is fully justified to
treat the next shell as a probe in the background generated by the
rest. The instability on this probe when the bound \eqn{JRQ} is
violated signals that this procedure must be physically forbidden
beyond the bound \eqn{JRQ}.

\section{Discussion}

We have found the exact IIA supergravity solution corresponding to
a source provided by the D2-brane supertube found in \cite{supertube}. We
have also found multi-tube solutions, which shows that parallel
supertubes exert no force on each other, and we have confirmed this by
showing that when (suitably aligned) stationary D0-brane, IIA-string and
D2-supertube probes are introduced into the background, they preserve all
supersymmetries of the background. This provides confirmation from
supergravity of the matrix model results of \cite{koreans} for
multi-supertubes. 

The supertube spacetime reproduces with remarkable accuracy the
worldvolume analysis of \cite{supertube}. It carries the appropriate
charges as measured by surface integrals at infinity: string charge
$Q_s$, D0 charge per unit length $Q_0$ and angular momentum $J$.
Its tension is $\tau = |Q_s| + |Q_0|$, as in \cite{supertube},
and it also preserves the same eight supercharges
(1/4 of those of the IIA Minkowski vacuum).
It is singular exactly
on a cylindrical surface of radius $R$ (but is regular everywhere
else). The only modification relative to the analysis in
\cite{supertube} comes from considering the possibility that the
D2-brane winds $N$ times around the tube. Taking this into
account, all the formulas for the supertube are precisely reproduced.
The radius is related to
the angular momentum as
$|J|= N R^2$. 
Finally, the requirement that there are no closed timelike curves in
the supertube spacetime imposes the bound $|J| \leq |Q_s Q_0| / N$
on the angular momentum, which is the same as in \cite{supertube},
once multiwinding is allowed for.

If $|J|$ exceeds the bound above then the Killing
vector field $\ell$, associated to rotations in the plane of the
angular momentum, becomes timelike sufficiently close to the tube
region, which leads to a global violation of causality.
Although this has no effect on the {\it local} physics of
D0-brane or IIA string probes, it causes an instability of
D2-supertube probes due to the fact that the supertube tension no
longer minimizes the energy for fixed D0 and string charges.
This is made possible by the appearance of a negative-norm state
(a ghost) on a supertube in the region where $\ell$ becomes timelike.
The global violation of causality that this produces (due to the
occurrence of CTCs) is thus manifested by a {\it local} pathology on
supertube probes (which is possible because these probes are
themselves non-local). This constitutes evidence that a globally
causality-violating supertube spacetime cannot be physically
assembled by starting with flat space and continuously bringing in from
infinity `supertube shells' with infinitesimal fractions of charge
and angular momentum.

It is straightforward to modify the IIA supergravity supertube solution
to one that provides the fields for a supertube in a Kaluza-Klein vacuum
spacetime of the form $\bE^{(1,n)}\times \mathcal{M}_{9-n}$, $n\geq 4$,
with $\mathcal{M}_{9-n}$ a compact Ricci-flat manifold. The metric then
takes the form
\bea
ds^2_{\it 10} &=& -U^{-1}V^{-1/2}(dt-A)^2+U^{-1}V^{1/2}dx^2\nonumber\\
&& +V^{1/2}[dr^2+r^2d\varphi^2+d\rho^2+\rho^2d\Omega_{n-4}^2
+ds^2(\mathcal{M}_{9-n})]\,.
\eea
For even $n$ the solutions involve elliptic functions, and therefore are
somewhat awkward to work with. For odd $n$, instead, they take simple
forms. For $n=7$ (a supertube in eight dimensions),
\bea
U&=& 1 + {|Q_s| \over 4\Omega_{\it 5}}{r^2+\rho^2+R^2\over
\Sigma^3}\,,\nonumber\\
A&=& {J\over 2\Omega_{\it 5}}{r^2\over \Sigma^3}\, d\varphi\,,
\eea
while for $n=5$,
\bea
U&=& 1 + {|Q_s| \over 2\Omega_{\it 3}\Sigma}\,,\nonumber\\
A&=& {J\over 2\Omega_{\it 3}}{r^2\over \Sigma (r^2+\rho^2+R^2+\Sigma)}\,
d\varphi\,.
\eea
$V$ is as $U$ with $Q_s$ replaced by $Q_0$, and $\Sigma$ is again
given by \eqn{defsigma}. In all cases it is possible to verify that
the absence of CTCs implies the bound \eqn{JRQ}.

The maximally rotating six-dimensional ($n=5$) solution is in fact dual to
the
solution for a helical D-string constructed in \cite{recent2}. One of
the two methods applied in \cite{recent2} to the construction of this
solution (the one based on the chiral null model) is essentially
equivalent to the one employed in this paper. The other approach, which
starts from the neutral rotating black hole and subjects it to several
transformations, does not lead to supertubes in dimensions higher than
six. Instead, it can be seen to result in filled-in cylinders, that is,
continuous distributions of concentrical tubes inside a cylinder. These
solutions do not possess CTCs either provided that $J$ does not exceed
the bound.

This six-dimensional supertube is particularly interesting when the
compact four-dimensional space is $K3$. Although at weak string coupling
the supertube source is distributional, and hence singular, one might
expect it to be non-singular at strong coupling. To examine this
possibility we should look for solutions of the six-dimensional
\mbox{supergravity/Yang-Mills} theory that governs the low-energy limit of
the
\mbox{$T^4$-compactified} S-dual heterotic string theory. The
distributional
D2-brane of the IIA theory now appears as a non-singular 2-brane with a
magnetic monopole core. Conceivably, a tubelike configuration of this
monopole 2-brane can be supported against collapse by angular momentum in
just such a way that its effective worldvolume description is as a $D$=6
supertube (that is, a d2-brane supertube of iia `little' string theory).
In this case, one might hope to find a `full' non-singular
supergravity/SYM solution that reproduces the fields
of the dualized IIA supergravity tube (after the redefinition of fields
required by the IIA/heterotic duality) but completed in the interior by
a solution with the `SYM-supertube' source. The case of the heterotic
dyonic instanton \cite{ETZ} provides a model for this kind of non-singular
completion of a rotating brane solution of supergravity, although in
that case the angular momentum is fixed by the charges rather than
just being bounded by them. 

In view of these connections, it appears
that the supergravity supertube spacetime studied here will eventually
take its place in a more general theory of supergravity solutions for
supersymmetric sources supported by angular momentum.

\acknowledgments

We thank Jerome Gauntlett and Carlos Herdeiro for discussions. R.E.
acknowledges partial support from UPV grant 063.310-EB187/98 and CICYT
AEN99-0315. D.M. is supported by a PPARC fellowship.

\newcommand{\NP}[1]{Nucl.\ Phys.\ {\bf #1}}
\newcommand{\AP}[1]{Ann.\ Phys.\ {\bf #1}}
\newcommand{\PL}[1]{Phys.\ Lett.\ {\bf #1}}
\newcommand{\CQG}[1]{Class. Quant. Gravity {\bf #1}}
\newcommand{\CMP}[1]{Comm.\ Math.\ Phys.\ {\bf #1}}
\newcommand{\PR}[1]{Phys.\ Rev.\ {\bf #1}}
\newcommand{\PRL}[1]{Phys.\ Rev.\ Lett.\ {\bf #1}}
\newcommand{\PRE}[1]{Phys.\ Rep.\ {\bf #1}}
\newcommand{\PTP}[1]{Prog.\ Theor.\ Phys.\ {\bf #1}}
\newcommand{\PTPS}[1]{Prog.\ Theor.\ Phys.\ Suppl.\ {\bf #1}}
\newcommand{\MPL}[1]{Mod.\ Phys.\ Lett.\ {\bf #1}}
\newcommand{\IJMP}[1]{Int.\ Jour.\ Mod.\ Phys.\ {\bf #1}}
\newcommand{\JHEP}[1]{J.\ High\ Energy\ Phys.\ {\bf #1}}
\newcommand{\JP}[1]{Jour.\ Phys.\ {\bf #1}}

\end{document}